\documentclass{article}
\usepackage{spconf,amsmath,graphicx}

\usepackage{url,amsfonts,amssymb,bm}

\usepackage{enumitem}
\usepackage{algorithm}
\usepackage{algorithmicx, algpseudocode }
\usepackage{multirow}
\usepackage{color}
\usepackage{mathtools}

\usepackage{caption}
\usepackage{subfigure}
\usepackage{amsthm}

\usepackage[colorlinks=true, linkcolor=blue, citecolor=blue, urlcolor=black]{hyperref}
\hypersetup{pdfcreator  = {LaTeX with hyperref package},
               pdfproducer = {dvips + ps2pdf} }

\newtheorem{theorem}{Theorem}

\newtheorem{proposition}{Proposition}

\theoremstyle{definition}

\DeclareMathOperator*{\argmin}{arg\,min}

\DeclareMathOperator*{\tr}{tr}
\DeclareMathOperator*{\diag}{diag}

\usepackage{bm}

\long\def\comment#1{}

\newfont{\bbb}{msbm10 scaled 700}

\newfont{\bb}{msbm10 scaled 1100}

\newcommand{\Ec}{{\cal E}}

\newcommand{\Gc}{{\cal G}}

\newcommand{\Tc}{{\cal T}}

\newcommand{\Vc}{{\cal V}}

\title{An efficient algorithm for graph Laplacian optimization \\ based on effective resistances}
\name{Eduardo Pavez and Antonio Ortega\thanks{This work was funded in part by NSF under grant CCF-1410009, and by a Google Faculty Research Award}}
\address{University of Southern California, Los Angeles, California, USA}
\begin{document}
\ninept
\maketitle
\begin{abstract}
In graph signal processing, data samples are associated to vertices 
on a graph, while  edge weights represent similarities between those samples. 
We propose a convex optimization problem  to learn sparse well connected graphs from data. We prove that each edge weight in our solution is upper bounded by the inverse of the distance between data features of the corresponding nodes. We also show that  the effective resistance distance between nodes is upper bounded by the   distance  between nodal data features. Thus, our proposed method learns a sparse well connected graph that encodes geometric properties of the data. 
We also propose a coordinate minimization  algorithm that, at each iteration, updates an edge weight using exact minimization.   The algorithm has a simple  and low complexity implementation based on closed form expressions.
\end{abstract}
\begin{keywords}
effective resistance, graph learning, graph Laplacian, sparsity, matrix tree theorem
\end{keywords}
\section{Introduction}
\label{sec:intro}
%
%
%
Graphs are  versatile tools for modeling high dimensional data and have been applied to data processing problems in various fields  \cite{garcia2008some, ishizaki2018graph,rue_gaussian_2005,ortega2018graph}. Graph  nodes represent  variables or features, and  edges represent dependencies between them.  
%
%
%
%
%
%
%
%
In this paper we are interested in learning, from data, graphs that are both {\em sparse} and {\em well connected}. 
Sparse graphs, i.e., those where the number of edges is in the order of the number of nodes, are desirable because they lead to graph based algorithms, e.g., graph filters \cite{ortega2018graph}, that have lower computational complexity and are easier to interpret. 
However, sparse graphs are more likely to be disconnected, which can be undesirable,  since  well connected graphs lead to graph algorithms that have higher performance, robustness and adapt better  to local geometry \cite{ren2005survey,olfati2004consensus,thrun2006graph,khosoussi2019reliable,yang2018designing,spielman2011graph,batson2013spectral}. 

There are various ways of quantifying how well connected a graph is. Some authors define well connected graphs as having small total resistance  \cite{yang2018designing,ghosh2008minimizing}, while others have defined them as having large algebraic connectivity \cite{ghosh2006growing}. These quantities have been used to characterize  the rate of information propagation  \cite{ren2005survey,olfati2004consensus}, and  communication among agents in a network \cite{thrun2006graph,khosoussi2019reliable}.
A third approach  is  based on  the matrix-tree theorem, which states that the number of spanning trees of a graph is equal to a  minor (determinant of a certain sub-matrix) of the Laplacian. Under these metrics,  connectedness can be improved by adding edges or by increasing edge weights. 
 
Graph learning algorithms select a graph topology (edge set) and edge weights to capture  similarity between data points (features) associated to nodes. Conventional data driven methods compute pairwise distances between nodes, based on their respective features, and select a graph topology based on these distances, e.g., by connecting nodes to their $K$ nearest neighbors. Edge weights  are assigned inversely proportional to the distance, e.g., using a  kernel similarity. 
 
 We take a more general approach and compute  a  cost  $h_{ij}$   which can be any  dissimilarity (distance-like function) between features at nodes $i$  and $j$, e.g., the distance  used in traditional approaches. The goal of these costs is to  encourage smaller (or zero) weights, when nodes in the graph are expected to be dissimilar. 
  Our graph learning algorithm follows two steps. 
 First we choose a sparse graph topology based on the edge costs, e.g., by  inputting them to a $K$ nearest neighbors algorithm.  We provide a theoretical justification for distance based graph topology selection in Section  \ref{sec_properties}. Second, we optimize the edge weights  while guaranteeing that the resulting graph remains well connected. 
 This is done by using a convex  negative log determinant objective function, for which  minimization is equivalent to maximizing the Laplacian minor (Section \ref{sec_algo}). 
 
Early work to learn sparse well connected graphs jointly optimized the topology and edge weights without using data \cite{ghosh2006growing,ghosh2008minimizing}. More recently, \cite{li_spanningTrees_2019_TIT} and others (see references therein), start from a weighted complete graph, and maximize graph connectivity (using the Laplacian minor) under sparsity constraints.   These approaches optimize the graph topology, and although the initial edge weights can be data dependent, they are not changed (optimized) by the algorithm.
 Others \cite{dong_learning_2016,kalofolias_how_2016,daitch2009fitting,berger2018graph} have proposed smoothness based graph learning problems, that lead to data adaptive sparse graphs. If the optimization problem only included a smoothness objective function, the optimal solution would be a graph  with all edge weights equal to zero.  To avoid these trivial solutions, additional  constraints and regularization terms that penalize graphs with isolated nodes and small  weights can be introduced \cite{dong_learning_2016,kalofolias_how_2016,daitch2009fitting,berger2018graph}. However,  the  resulting graphs might still have  disconnected components, and choosing a good set of  parameters that balances sparsity, connectedness and  data fidelity becomes an additional burden. 
 
Our first contribution is a theoretical analysis  of the case when  the input graph topology is fixed but otherwise arbitrary (e.g., it could be the complete graph).  We show that  for a particular choice of edge costs (see Section \ref{sec:prel}), the proposed optimization problem is equivalent to estimation of attractive intrinsic Gaussian Markov random fields (GMRF)  \cite{rue_gaussian_2005}. This problem was initially studied in \cite{egilmez2017graph} as an extension of the graphical Lasso  \cite{friedman_sparse_2008}, where the precision matrix is the  Laplacian.   We prove two important properties of this optimal graph that are related to connectedness and sparsity (Proposition \ref{prop_solution_bounds}). First, we  show that the effective resistance distance \cite{ellens2011effective} of an edge  is upper bounded by the edge cost, thus enforcing that similar nodal features (small cost) are mapped to nearby graph nodes. Second, the magnitude of the optimal edge weight is bounded by the inverse of  the edge cost, thus ensuring far apart data features, are related through a small edge weight in the optimal graph. Although several papers have studied this solution from  probabilistic and algorithmic perspectives \cite{egilmez2017graph,hassan2016topology,zhao2019optimization,egilmez2018},  the derived bounds are new and provide  insights into the geometric properties of the solution. In particular, these properties justify the use of   distance based  graph topology selection algorithms (e.g., $K$  nearest neighbors)   when optimizing the Laplacian minor.

Our second contribution is a coordinate minimization algorithm to learn the edge weights. 
We obtain a closed form expression for each update that depends on the edge's effective resistance and cost.
By exploiting some results  on learning trees \cite{lu2018learning}, we show that the proposed method can be implemented with $\mathcal{O}(n^2)$ complexity per iteration. The complexity  is  lower because  we avoid  a naive implementation that takes $\mathcal{O}(n^3)$ time per iteration. Compared to the first algorithm for this problem \cite{egilmez2017graph}, and more recent improvements and extensions \cite{hassan2016topology,zhao2019optimization,egilmez2018}, our algorithm has lower complexity per iteration. In addition, once the data (in the form of edge costs),   edge set $\mathcal{E}$,  and a convergence tolerance are chosen, the proposed algorithm has no other parameters. Our code is available online\footnote{\url{https://github.com/STAC-USC/graph_learning_CombLap}}.

The rest of this paper is organized as follows.  In Section \ref{sec:prel} we introduce the graph learning problem and some graph theoretic concepts. Section \ref{sec_properties} is concerned with theoretical  properties of the optimal graph Laplacian. Section \ref{sec_algo} introduces the proposed algorithm, while Section \ref{sec_exp} provides numerical evaluation. Section \ref{sec_conclusion} provides some conclusions and future research directions.
%
%
%
%
%
\section{Preliminaries}
\label{sec:prel}
\subsection{Graph theoretic background}
We  denote scalars, vectors and matrices using lowercase regular, lowercase bold, and uppercase bold font respectively, e.g., $a$,  $\mathbf{a}$, and $\mathbf{A}$. The symbols $\dagger$ and $^{\top}$ indicate  Moore-Penrose pseudo inverse and transpose respectively. The all ones $n\times n$ matrix is denoted by $\mathbf{J}_n$.
We consider undirected connected graphs with positive weights and without self loops. 
A graph $\Gc=(\Vc,\Ec, \mathbf{W})$ with node set $\mathcal{V} = \lbrace 1,\cdots, n \rbrace$, edge set $\mathcal{E} \subset \mathcal{V} \times \mathcal{V}$ has individual edges denoted by $e = (i,j) \in \mathcal{E}$, with corresponding edge weight $w_e = w_{ij}$. The weight matrix $\mathbf{W} = (w_{ij})$ is symmetric and  non negative, and  $w_{ij} > 0$ if and only if $  (i,j) \in \mathcal{E}$ or $(j,i) \in \mathcal{E}$.  Since we consider undirected graphs,  we do not repeat edges when listing them, that is,  if $(i,j) \in \mathcal{E}$ we do not include $(j,i)$ in the edge set.
We also denote by $\mathbf{w}$, the $m$ dimensional vector with entries $w_{e}$ for $e \in \Ec$. The degree   matrix is $\mathbf{D} = \diag(\mathbf{W 1})$. 
For an edge $e=(i,j)$, the edge vector is denoted by $\mathbf{g}_e \in \mathbb{R}^n$, which has entries $\mathbf{g}_e(i) =1$,  $\mathbf{g}_e(j) = -1$, and zero otherwise.  The incidence matrix is $\mathbf{\Xi} = [\mathbf{g}_{e_1}, \mathbf{g}_{e_2}, \cdots, \mathbf{g}_{e_m}]$, while the combinatorial Laplacian matrix is defined as
$ \mathbf{L} = \mathbf{D -W} = \mathbf{\Xi \diag( w) \Xi^{\top}}$.
 The strengthened Laplacian is defined as $ \mathbf{Q} = \mathbf{L} + \frac{1}{n}\mathbf{J}_n $ \cite{hassan2016topology}.
 The set of combinatorial Laplacian matrices of graphs   with $n$ nodes is denoted by $\mathbb{L}_n$.
 The eigenvalues of $\mathbf{L}$ are denoted by $0=\lambda_1 \leq \lambda_2 \leq ,\cdots,\leq  \lambda_n$. 
%
To determine when a graph is well connected we define:
\begin{itemize}
\item A \emph{path} from node $i$ to node $j$ is a sequence of edges $\mathcal{P}_{ij}=\lbrace (i_1, i_2),  \cdots, (i_{t-1}, i_t) \rbrace \subset \mathcal{E}$ so that $i = i_1$, and $j =i_t$.
\item A graph  is \emph{connected} if  $\forall (i,j) \in \mathcal{V} \times \mathcal{V}$, there exists $\mathcal{P}_{ij} \neq \emptyset$. A graph is connected if and only if $\lambda_2>0$, or if $\mathbf{Q}\succ 0$.
\item A  \emph{tree}  is a connected  graph with $n-1$ edges.
\item A \emph{spanning tree} of  $\mathcal{G}=(\Vc,\Ec,\mathbf{W})$ is any sub-graph with vertex set $\Vc$ that is a tree.
\end{itemize}
%
%
%
%
The weight of a spanning tree  $\mathcal{T} \subset \Ec$ of  graph  $\mathcal{G}=(\Vc, \Ec, \mathbf{W})$ is the product of its edge weights, that is,
$\Omega(\mathbf{L}_\mathcal{T}) = \prod_{e \in \mathcal{T}}w_e$, 
where $\mathbf{L}_{\Tc} = \sum_{e \in \Tc} w_e \mathbf{g}_e \mathbf{g}_e^{\top }$ is the Laplacian of the tree. 
Define the sum of weights of all spanning trees of $\Gc$ with Laplacian $\mathbf{L}$ as:  
\begin{equation}\label{eq_num_spanning_trees}
    \Omega(\mathbf{L}) = \sum_{\mathcal{T} \subset \Ec}\Omega(\mathbf{L}_\mathcal{T}). 
\end{equation}
Note that for unweighted graphs (all $w_e=1$) all trees have unit weight, hence  $\Omega(\mathbf{L})$ is the  number of spanning trees. Also note that for disconnected graphs $\Omega(\mathbf{L})$ is  zero. 
%
Graphs with larger $\Omega(\mathbf{L})$  have  more spanning trees of larger weights, so we can use this quantity to quantify how well connected a graph is.
Based on (\ref{eq_num_spanning_trees}) a graph connectivity can increase  by adding edges (more spanning trees) or by increasing weights for existing edges (of existing spanning trees).  Computing  $\Omega(\mathbf{L})$ from its definition would require enumerating all spanning trees, fortunately there is an elegant alternative. 
   \begin{theorem}[Matrix tree theorem]  $\Omega(\mathbf{L})$ 
can be computed as:
\begin{equation}\label{eq_matrix_Tree}
   \Omega(\mathbf{L}) = \frac{1}{n}\det\left(  \mathbf{L} + \frac{1}{n}\mathbf{J}_n\right) = \frac{1}{n}\prod_{i=2}^n \lambda_i.
\end{equation}
\end{theorem}
  For any pair of vertices $i$ and $j$, their effective resistance   is:   
	\begin{equation*}
	r_{ij}=r_e = \mathbf{g}_{e}^{\top} \mathbf{L}^{\dagger} \mathbf{g}_e = \mathbf{g}_{e}^{\top} \mathbf{Q}^{-1} \mathbf{g}_e,
	\end{equation*}
where  $e = (i,j) \in \Vc \times \Vc$. 
  $r_{ij}$   obeys the triangular inequality and other metric properties \cite{ellens2011effective}, thus nodes $i$ and $j$ are well connected   when $r_{ij}$ is small.
Since $\log\det( \mathbf{Q} ) = \log(n \Omega(\mathbf{L}))$, we have that
\begin{equation*}
\frac{\partial \log\det( \mathbf{Q} )}{\partial w_e}  =\tr\left(\mathbf{Q}^{-1} \frac{\partial \mathbf{Q}}{\partial w_e}\right)
=\tr\left(\mathbf{Q}^{-1} \mathbf{g}_e\mathbf{g}_e^{\top}\right) = r_e.		
\end{equation*}
\subsection{Edge cost}
 Data is available in matrix  $\mathbf{X} \in \mathbb{R}^{n \times N}$.  $\mathbf{x}^{(k)}$ is the $k$-th row of  $\mathbf{X}$, and corresponds  to  the data point (feature) associated to the $k$-th node. Alternatively, the $k$-th column of $\mathbf{X}$ is denoted by  $\mathbf{x}_k$, which can be viewed as a graph signal. We define $h_e$ as the quantity that represents the cost of including  edge weight $e=(i,j)$ in the graph. Intuitively, edges with small costs are more likely to be given larger weights.
  Examples of edge costs are given next.
 
 {\bf Estimation of attractive intrinsic Gaussian Markov random fields (GMRF).} If the edge cost is chosen as
 \begin{equation}\label{eq_cost_gmrf}
      h_{e}=h_{ij} = \alpha + \frac{1}{N}\sum_{k=1}^N ( x_{ik} - x_{jk}  )^2,
 \end{equation}
 then $\sum_{e} h_e w_e =\tr(\mathbf{L S}) + \alpha \Vert \mathbf{w} \Vert_1$,  where $\mathbf{S} = \frac{1}{N} \mathbf{X} \mathbf{X}^{\top }$.  This cost term appears  in  estimation of  sparse GMRFs that have  a precision matrix in the form of a Laplacian  \cite{egilmez2017graph,hassan2016topology, zhao2019optimization}, under the assumption that $\mathbf{x}_k$ are zero mean independent identically distributed  graph signals.

{\bf Learning graphs from smooth signals.} For any  $p>0$, the $\ell_p$ Laplacian variation of a signal $\mathbf{x}$ is $ V_p(\mathbf{x}) = \sum_{(i,j) \in \Ec }w_{ij} \vert x_i -x_j \vert^p$.
   The average variation of the columns of  $\mathbf{X}$ is
\begin{equation}\label{eq_cost_average_Variation}
    \frac{1}{N}\sum_{k=1}^N V_p(\mathbf{x}_k) = \sum_{(i,j) \in \Ec }w_{ij} {\frac{1}{N}\sum_{k=1}^N   \vert x_{ik} -x_{jk} \vert^p} = \sum_{(i,j) \in \Ec} w_{ij} h_{ij}.
\end{equation}{}
 Recently, \cite{dong_learning_2016,kalofolias_how_2016,berger2018graph} used this criterion to learn sparse graphs.

{\bf Similarity graph optimization.} Note that the costs in  (\ref{eq_cost_gmrf}) and (\ref{eq_cost_average_Variation}) are distances between the rows of $\mathbf{X}$.  We can choose any distance-like function, for example a Gaussian kernel  cost  
\begin{equation}\label{eq_gaussian_dist}
h_{ij } = \exp\left({\Vert \mathbf{x}^{(i)} - \mathbf{x}^{(j)} \Vert^2}/{\sigma^2}\right).
\end{equation}
\section{Problem formulation and  properties}
\label{sec_properties}
In this section we state the graph learning problem and establish some properties of the solution. We divide the problem into two steps: edge set selection, and edge weight estimation.  Edge weights are obtained by solving the optimization problem 
\begin{equation}\label{eq_prob_log_formulation}
\min_{\mathbf{L} \in \mathbb{L}_n}  -\log \det(\mathbf{Q})
\textnormal{ s.t.}, \quad w_e =0,~ e \notin \mathcal{E} \quad \sum_{e \in \mathcal{E} } h_{e} w_e \leq C.
\end{equation}
 Note that  the negative log determinant is convex, and ensures the graph is connected.
Without loss of generality we pick  $C=n-1$.

In (\ref{eq_prob_log_formulation}),  we assume that the edge set $\Ec$ is given. If no prior information is available, $\Ec$ can be chosen as the complete graph.
Second, we address the problem of choosing an edge set $\Ec$ that is sparse, i.e., it has at most $m =\mathcal{O}(n)$ edges. Our approach is based on a theoretical analysis of the solution of (\ref{eq_prob_log_formulation}) when $\Ec$ is the complete graph. 
%
 The Lagrangian of (\ref{eq_prob_log_formulation}) is
\begin{equation*}
-\log\det(\mathbf{Q})  + \nu \left( \sum_{e\in \mathcal{E}} h_e w_e - (n-1) \right)  -  \sum_{e \in \mathcal{E}} \lambda_e w_e   -  \sum_{e \notin \mathcal{E}} \gamma_e w_e.
\end{equation*}{}
Then, the Karush Kuhn Tucker (KKT) optimality  conditions are, 
\begin{align*}
    &- r_e + \nu h_e - \lambda_e = 0,\quad 
    \lambda_e, w_e \geq 0, \quad \lambda_e w_e=0,\quad \forall e \in \mathcal{E}, \\
    &\sum_{e\in \mathcal{E}} h_e w_e \leq n-1, \quad 
    \nu(\sum_{e\in \mathcal{E}} h_e w_e -(n-1))=0,\quad     \nu \geq 0, \\
    &-r_e  - \gamma_e = 0, \quad \forall e \notin \mathcal{E}. 
\end{align*}{}
Any connected graph obeys    $\sum_{e} r_e w_e = n-1$, which    combined with the KKT conditions  implies that $\nu =1$ and $\sum_{e} w_e h_e = n-1$. After further algebraic manipulations we obtain the following.
\begin{proposition}\label{prop_solution_bounds}
	The effective resistances  of the solution of (\ref{eq_prob_log_formulation}) obey
	\begin{equation}
		r_e^* = h_e, \forall e \in \mathcal{E}^*,\quad
		r_e^* \leq h_e,  \forall e \in \mathcal{E}\setminus \mathcal{E}^*,
	\end{equation}
	while the weights satisfy
	\begin{equation}\label{eq_opt_weight_bound}
		w^*_e  \leq {1}/{h_e}, ~ \textnormal{ if }~ e \in \mathcal{E}^*.
	\end{equation}
\end{proposition}
These bounds  explicitly associate the  data similarity to the connectedness and sparsity of the graph through the effective resistance and graph weights, respectively. When features of nodes $i$ and $j$ are close, i.e., $h_{ij}$ is small,  the optimal edge will have a low effective resistance, thus ensuring nearby  nodes are well connected in the graph. 
On the other hand, when $h_{ij}$ is large, the corresponding optimal weight is close to zero.  Similar bounds based on correlation instead of distance  are  available  when the goal is learning a  generalized graph Laplacian  \cite[Ch. 4]{pavez_thesis2019}.

 An important consequence of Proposition \ref{prop_solution_bounds} is that it reveals information about the optimal graph, without having to solve (\ref{eq_prob_log_formulation}). Since an edge with large costs $h_{ij}$  will have zero or small   weights, a  sparse edge set $\Ec$ can be chosen  as the subset of the complete graph, for which the pairs $(i,j)$  have a small edge cost.  We illustrate this numerically in Section \ref{sec_exp}, using  $K$ nearest neighbor algorithm. 
%
 \section{Coordinate minimization algorithm}
\label{sec_algo}
 
It can be proven that   the solution of (\ref{eq_prob_log_formulation}) is equal to the solution of
\begin{equation}\label{eq_maxLike_problem1}
\min_{\mathbf{L} }F(\mathbf{L}) \quad \textnormal{ s.t. }~ w_e \geq 0, ~e\in \mathcal{E},\quad w_e =0,~ e \notin \mathcal{E},
\end{equation}
where $F(\mathbf{L}) = -\log\det( \mathbf{Q}) + \sum_{e} h_e w_e$.
In this section we derive an iterative algorithm for solving (\ref{eq_maxLike_problem1}), that at each step   solves
 \begin{equation}\label{eq_coord_min}
 w_e^{(t+1)} = \argmin_{w_e \geq 0} F(\mathbf{L})~\textnormal{ s.t. } w_f = w_f^{(t)}, \forall f \neq e,
 \end{equation}
 for $e \in \mathcal{E}$.
 This type of algorithm  is known as \emph{coordinate minimization}. For general convex optimization problems,  one can solve  (\ref{eq_coord_min}) using coordinate descent with line search, which can be computationally expensive since it requires multiple evaluations of  the objective function. For this problem, however, there is a closed form solution.  Let $\mathbf{L}^{(t)}$, and  $\mathbf{Q}^{(t)}$ be the estimates at the $t$-th iteration, so that $\mathbf{L}^{(t)} + (1/n)\mathbf{J}_n = \mathbf{Q}^{(t)}$. The  update of the graph Laplacian is
\begin{equation}\label{eq_L_update}
\mathbf{L}^{(t+1)} =  \mathbf{L}^{(t)}  + \delta_e^{(t+1)} \mathbf{g}_e \mathbf{g}_e^{\top}.
\end{equation} 
 where $\delta_e^{(t+1)} = w_e^{(t+1)} - w_e^{(t)}$.  Our main result is stated below.
\begin{theorem}\label{th_optimal_weights}
  If  $\mathbf{L}^{(t)}$ is irreducible, then  (\ref{eq_coord_min}) is solved by
   \begin{equation}\label{eq_we_update}
 w_e^{(t+1)} = \max \left(0, w_e^{(t)} + {1}/{h_e} - {1}/{r_e^{(t)}}   \right), 
 \end{equation}
 and the updated  Laplacian $\mathbf{L}^{(t+1)}$ is also irreducible.
 \begin{proof}
  The update (\ref{eq_L_update}) changes the determinant as follows
\begin{equation}\label{eq_determinant_update}
 \det(\mathbf{Q}^{(t)} + \delta_e^{(t+1)} \mathbf{g}_e \mathbf{g}_e^{\top}) 
 =  (1 + \delta_e^{(t+1)}r_e^{(t)})\det(\mathbf{Q}^{(t)}).
 \end{equation}
 The change in the objective function is
 \begin{align}\label{eq_delta_F}
 F(\mathbf{L}^{(t+1)}) - F(\mathbf{L}^{(t)}) = -\log(1 + \delta_e^{(t+1)}r_e^{(t)}) + \delta_e^{(t+1)}h_e.
 \end{align}
  (\ref{eq_coord_min}) can be solved by minimizing (\ref{eq_delta_F}) subject to $w_e \geq 0$. The optimality conditions of that problem are
 \begin{equation*}
 \frac{-r_e^{(t)}}{1 + \delta_e^{(t+1)}r_e^{(t)}} + h_e -\lambda_e =0, ~ \lambda_e, w_e^{(t+1)} \geq 0, ~ \lambda_e w_e^{(t+1)} =0,
 \end{equation*}
 which are satisfied by (\ref{eq_we_update}).
  Note that $\mathbf{L}^{(t+1)}$ is irreducible if and only if the corresponding graph associated to its non zero pattern is connected. This is equivalent to    $\mathbf{Q}^{(t+1)} \succ 0$.  Since $\mathbf{Q}^{(t+1)}$ is positive semi-definite by construction,  we only have  to prove that it is also non-singular. Using (\ref{eq_determinant_update}), and since we assume that $\det(\mathbf{Q}^{(t)})> 0$, we  need to show that $1+ r_e^{(t)} \delta_e^{(t+1)}>0$.  (\ref{eq_we_update}) implies  $\delta_e^{(t+1)}  \geq   {1}/{h_e} -{1}/{r_e^{(t)}}$,  which combined with    $r_e^{(t)}>0$, and   $h_e>0$, gives the desired result.
 \end{proof}
 \end{theorem} 
  \begin{figure}[t]
     \centering
     \includegraphics[width=0.4\textwidth]{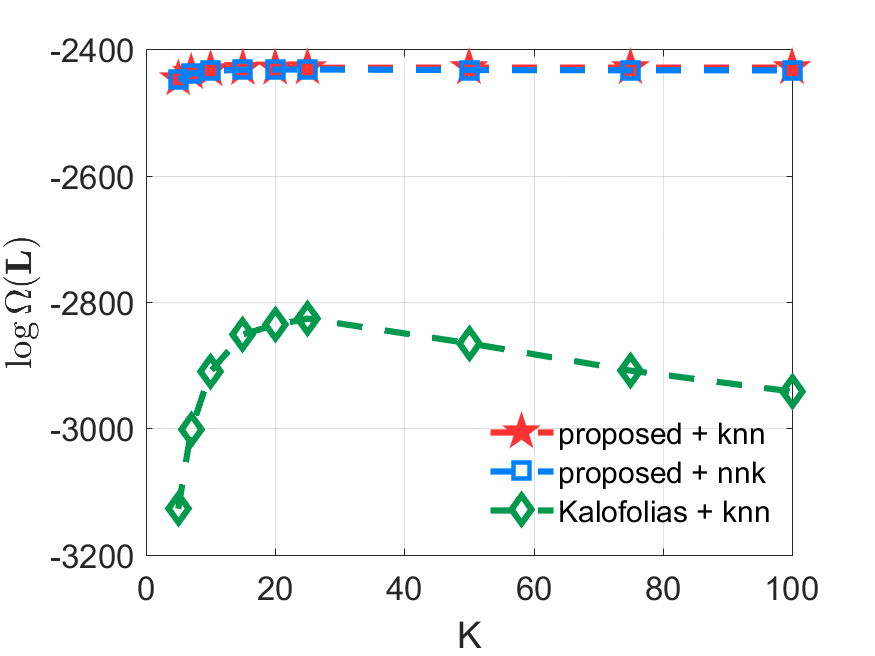}
     \caption{Effect of graph topology inference on  connectedness.}
     \label{fig:vs_kalo_knn_nnk}
 \end{figure}
    Coordinate minimization algorithm based on Theorem \ref{th_optimal_weights} can be implemented in $\mathcal{O}(n^2)$ time per edge weight update. Complexity is dominated by computation of the effective resistance. This and other issues related to computational complexity  are  discussed next.

{\bf Initialization.}
Theorem \ref{th_optimal_weights} is useful if we can find an initial estimate  $\mathbf{L}^{(0)}$ that is irreducible, that is, its graph is connected and contained in $\mathcal{E}$.
To find  $\mathbf{L}^{(0)}$   we use the graph  Laplacian that minimizes $F(\mathbf{L})$ over all spanning trees   $\mathcal{T} \subset \mathcal{E}$. This problem was solved in \cite{lu2018learning}, with complexity $\mathcal{O}(\vert \Ec \vert \log(n))$.  

{\bf Computing effective resistances.}
All effective resistances of a tree can be computed in $\mathcal{O}(n^2)$ time using Gaussian elimination with a  perfect elimination ordering \cite{vandenberghe2015chordal}. For $t>0$, the matrix $(\mathbf{Q}^{(t+1)})^{-1}$ can be  updated using  the Sherman-Morrison formula
\begin{equation}\label{eq_Sigma_update}
(\mathbf{Q}^{(t+1)})^{-1}=\mathbf{\Sigma}^{(t+1)} = \mathbf{\Sigma}^{(t)} - \frac{\delta_e^{(t+1)}(\mathbf{\Sigma}^{(t)} \mathbf{g}_e)(\mathbf{\Sigma}^{(t)} \mathbf{g}_e)^{\top}}{1 + \delta_e^{(t+1)}\mathbf{g}_e^{\top}\mathbf{\Sigma}^{(t)} \mathbf{g}_e }.
\end{equation}
Given  $e =(i,j)$ and $f=(s,t)$,  the updated effective resistances is
\begin{equation}
r_{f}^{(t+1)} = r_{f}^{(t)} - \left(\delta_e^{(t+1)} z_{ef}^{(t)}\right)/\left(1 + \delta_e^{(t+1)}r_e^{(t)} \right), 
\end{equation}
where   $z^{(t)}_{ef} =  ( {-r^{(t)}_{is} + r^{(t)}_{it} + r^{(t)}_{js} - r^{(t)}_{jt}}  )^2/4$.  By keeping the matrix $\mathbf{\Sigma}^{(t)}$ in memory,  all effective resistances can be computed in $\mathcal{O}(n^2)$ time after each weight update.
%
 
{\bf Edge selection rules.} 
In theory,   various rules  lead to algorithms with the same convergence rate. However, greedy rules have been shown to provide great improvements for some cases \cite{nutini2015coordinate}. We implement  cyclic, random and a greedy  rule. The {\em cyclic} rule iterates over edges $e \in \Ec$ in a fixed order. The {\em randomized} rule picks edges $e \in \Ec$ uniformly at random.   Both  have complexity $\mathcal{O}(1)$.
{\em Greedy} rules pick the edge that maximizes a certain criteria.  We use the  {\em proximal gradient Gauss-Southwell  (PGS)} rule, which  chooses the edge with the   largest  $\Delta_e^{(t)} = \vert \delta_e^{(t)} \vert$.
The PGS rule has complexity $\mathcal{O}(\vert \Ec \vert )$ per iteration.
 
 {\bf Convergence.} We say the coordinate minimization algorithm has converged if the decrease of the objective function after  one epoch ($\vert \Ec \vert$ iterations) is below a predefined threshold.  There is no need to check feasibility, since the iterates are guaranteed by Theorem \ref{th_optimal_weights} to be connected.  The initial value of the objective function $F(\mathbf{L}^{(0)})$ can be computed in $\mathcal{O}(n)$ time \cite{lu2018learning}. For $t\geq 1$, we update the objective function using (\ref{eq_delta_F}) in $\mathcal{O}(1)$ time, therefore there is no additional computational burden to check feasibility or convergence.
\section{Experiments}
\label{sec_exp}
To evaluate our theoretical results we learn graphs for the USPS handwritten digits \cite{lecun1990handwritten}. We randomly select $100$ images from each digit, to form a $1000$  data set, and learn a  $n=1000$ nodes similarity graph. Edge costs are computed using  (\ref{eq_gaussian_dist}).

\begin{figure}[t]
\centering 
    \includegraphics[width=0.4\textwidth]{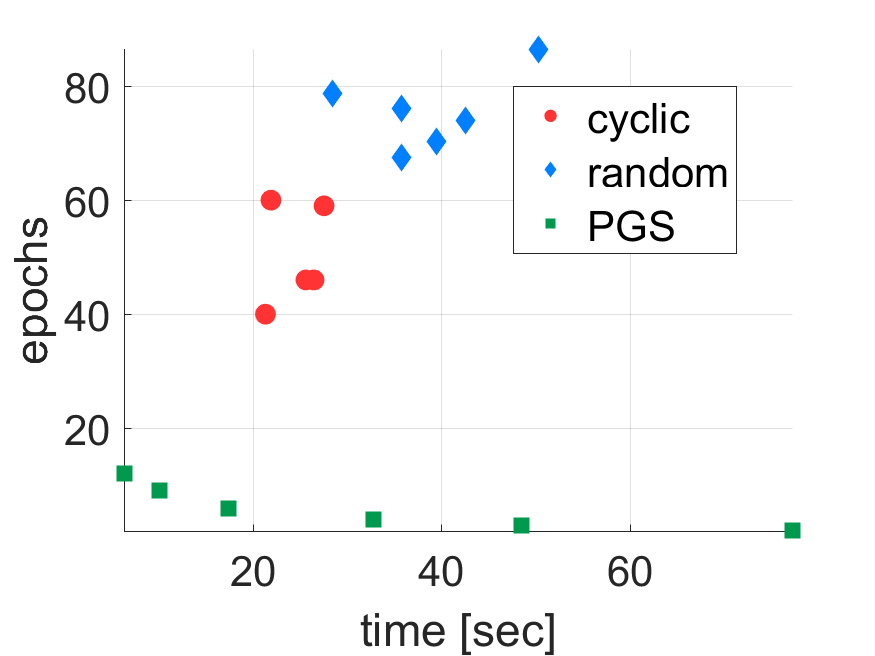}
    \caption{Effect of edge selection rules on  convergence.}
    \label{fig:epochs_vs_time}
\end{figure}

{\bf Graph topology inference. }
We study how  sparsity of the  input edge set $\Ec$ affects the learned graph. We  consider two methods for choosing the graph topology. We use  $K$ nearest neighbor graphs (KNN), where each node is connected to its $K$ closest points. We also use a method based on non negative kernel (NNK) regression \cite{shekkizhar2019graph}, which is a low complexity algorithm that sparsifies KNN graphs based on local geometry. We compute the logarithm of (\ref{eq_num_spanning_trees}) for the solution of (\ref{eq_prob_log_formulation})  as a function of number of nearest neighbors $K$. The input graph topology $\Ec$ is obtained using the KNN and NNK algorithms. We  compare against \cite{kalofolias_how_2016}, since that algorithm also uses as input a KNN graph topology.  Figure \ref{fig:vs_kalo_knn_nnk} indicates that the optimal graph Laplacian does not change when the input topology gets more dense (by increasing $K$). This can be explained by Proposition \ref{prop_solution_bounds}, which indicates that the magnitude of the optimal edge weights decays inversely proportional to the edge cost. The graph sparsity and density of edge weights becomes stable as $K$ increases. This indicates that in practice, a small value of $K$ can be chosen, which can have significant reduction on computational complexity (see Fig. \ref{fig:epochs_vs_time}). Another interesting observation is the fact that the NNK graph, even tough it is much sparser than the KNN graph (see \cite{shekkizhar2019graph}), produces a similar  graph Laplacians. In contrast, the graph learning method from \cite{kalofolias_how_2016} returns a graphs whose connectedness decreases when the input topology becomes more dense.    

{\bf Convergence and complexity  of edge selection rules.} We compare the cyclic, randomized and PGS edge selection for the same data set created for the previous experiment.  We say that a  algorithm has converged if after one epoch, the decrease in the  objective function is smaller than $10^{-10}$. We choose edge sets using KNN with $K \in \lbrace 5, 10, 20, 40, 60, 100 \rbrace$. The  number of epochs versus convergence time are plotted  in Figure \ref{fig:epochs_vs_time}. Using the cyclic rule lead to  convergence in fewer iterations than using random edge selection. Also, as predicted by \cite{nutini2015coordinate}, coordinate minimization with the PGS rule converges is much fewer iterations than with the randomized and cyclic rules. On the other hand, the complexity per iteration grows proportionally with the edge density, thus the PGS rule should be used for sparse graphs, while the cyclic rule for more dense graph topologies.
\section{Conclusion}
\label{sec_conclusion}
We have analyzed a framework for learning combinatorial graph Laplacian matrices from data, with  the goal of obtaining a sparse and well connected graph. Our theoretical analysis indicates that the optimal edge weights decay inversely proportional to the distance between nodal features, thus suggesting that distance like graph topology inference algorithms, such as KNN  and others \cite{shekkizhar2019graph} can be used for edge set selection. 
We also propose a coordinate minimization algorithm that has low complexity per iteration thus it  can be used   to learn sparse graphs with hundreds to thousand of nodes. 
%
%
\vfill
\pagebreak
\bibliographystyle{IEEEbib}
\bibliography{refs}
%
%
%
%
%



\end{document}